\documentclass{article}
\usepackage{graphicx} 

\usepackage[super,sort&compress,comma]{natbib} 
\usepackage[version=3]{mhchem}
\usepackage[left=2.5cm, right=2.5cm, top=1.785cm, bottom=2.0cm]{geometry}
\usepackage{balance}
\usepackage{mathptmx}
\usepackage{sectsty}
\usepackage{lastpage}
\usepackage[format=plain,justification=justified,singlelinecheck=false,font={stretch=1.125,small,sf},labelfont=bf,labelsep=space]{caption}
\usepackage{float}
\usepackage{fancyhdr}
\usepackage{fnpos}
\usepackage[english]{babel}
\addto{\captionsenglish}{%
}
\usepackage{array}
\usepackage{droidsans}
\usepackage{charter}
\usepackage[T1]{fontenc}
\usepackage[usenames,dvipsnames]{xcolor}
\usepackage{setspace}
\usepackage[compact]{titlesec}
\usepackage{hyperref}
\usepackage{siunitx}

\usepackage{epstopdf}

\title{Direct separation of intra- and inter-molecular contributions in pulse dipolar EPR experiments}
\author{Olga Vojtiskova$^{1}$, Laura Galazzo$^{1}$, Maxim Yulikov$^{1*}$}
\date{\today}

\begin{document}

\maketitle
$^{1}$ Department of Chemistry and Applied Biosciences, Vladimir-Prelog-Weg 2, 8093, Zurich, Switzerland\\[1em]
*E-mail: myulikov@ethz.ch

\section{Abstract}

We present a direct method of separating intra- (form factor) and inter-molecular (background) contributions in pulse EPR dipolar spectroscopy (PDS) experiments. The form factor/background separation is accomplished by rearranging data measured at two different biradical concentrations, without any fitting fundamentally required. This new data processing method allows to analyse PDS data from spin concentrated samples for which background separation was previously unrealistic. In particular, the presented method would be of high importance for studies of spin labeled biomolecules, especially in respect to the liquid-liquid phase separation (LLPS). Furthermore, we propose a simple measurement and data processing protocol to separate the PDS contributions from each phase in LLPS samples. The protocol for the form factor/background separation for RIDME experiment is also discussed.

\section{Introduction}

Pulse dipolar spectroscopy in combination with site-directed spin labelling has been established as a well accepted structural biology method,\cite{jeschke_annurev_2012,Bordignon2012} particularly appreciated in studies of biomolecules with intrinsic disorder.\cite{Masliah2018,Emmanouilidis2021,ritsch_acie_2022,Dorn2023,EstebanHofer2024,Nguyen2026} In PDS methodology, determination of conformational distribution of a biomolecule requires measurement of several site-to-site distance distributions. The accessible distance range varies to some extent between different types of PDS experiments, and it is fundamentally restricted by the spin label's phase memory timescale. The standard PDS sample preparation and measurement protocol requires preparing solution of spin labelled biomolecules at low enough bulk concentration, so that the expected intramolecular distances are substantially shorter than the characteristic intermolecular distances.\cite{Schiemann2021} Accordingly, a good quality PDS time trace of a biomolecule with broad conformational distribution contains, typically, an oscillations-free form factor signal that decays well before the end of the trace and a background signal that decays substantially slower than the characteristic decay in the form factor. As a result, such trace has a kink, where transition from substantial form factor contribution to almost exclusively background contribution takes place. In the simplest protocol, the background function is then fitted within the time range beyond the kink.\cite{Jeschke2006} Even if background and form factor are fitted together in a more advanced global fitting routine, as offered e.g. by DeerLab package,\cite{FabregasIbanez2020} still, a reliable background separation depends on the presence of such a kink region. It happens regularly that due to the unfortunate relations between maximum affordable length of the PDS time trace, the minimum concentration sensitivity limit, and the range of intramolecular distances to be detected, the characteristic decay rates of the form factor and background signals are similar, and the standard separation protocol reports on large uncertainties. To date, such cases were considered as uninterpretable.

One can, however, avoid this obstacle by measuring PDS data at two or more different bulk spin concentrations, because the intermolecular background contribution is concentration dependent, while the form factor signal does not change with the bulk concentration as long as the crowding effects can be neglected. Ideally, this should be done by spin dilution in a solution of the biomolecules pertaining identical total bulk biomolecules' concentration for all samples in the series. This would help bypassing any possible effects due to crowding and long-range protein interaction potentials. The intramolecular distance distribution would thus be independent on the bulk spin concentration, whereas the background signal will be changing according to the change of the spin concentration. Here, we describe the method of form factor and background separation that exploits the additional information obtained in such PDS measurements with varying spin concentrations. We also discuss differences in the form factor/background separation protocols for double electron-electron resonance (DEER)\cite{milov1981application,milov1984electron,pannier2011dead} and relaxation induced dipolar modulation enhancement (RIDME)\cite{kulik2001electron,milikisyants2009pulsed} as well as the method to separate PDS signals from the two phases in LLPS samples.


\section{Statistical considerations for the PDS signal}

For the constant-time DEER, as well as for the variable time DEER and the single frequency PDS experiments in fully deuterated samples, one can represent the PDS signal $V_s(t)$ before ensemble averaging in the form:

\begin{equation}
\label{eq:PDS_0}
    V_s(t) = F_s(t)\cdot\prod_{i=1}^{N}b_{s,i}(t).
\end{equation}

Here $F_s(t)$ is the intra-molecular form factor and $b_{s,i}(t)$ is the intermolecular PDS signal for the coupling of the detected spin ('A spin') and the spin (or spins) in a different molecule ('B spins'), indexed with 'i'. Label 's' indicates that we take one specific conformation, position and orientation with respect to the static magnetic field, i.e. it corresponds to the term before statistical ensemble averaging. In case of doubly spin labeled biomolecules and nitroxide spin labels (with spin $S=1/2$), using the inversion efficiency $\beta_{i.j}$ one can write: 

\begin{equation}
\label{eq:PDS_b}
    b_{s,i}(t) = \prod_{j=1,2}\left[(1-\beta_{i.j}) + \beta_{i,j}\cos(\omega_{i,j}t)\right].
\end{equation}

Clearly, this can be extended to any number of spin labels attached to each biomolecules. This equation can also be substituted by a weighted sum of different contributions, in case of labelling efficiency below 100\%. For high-spin centers used as spin labels, $b_{s,i}(t)$ as well as the form factor formula can also contain additional terms due to the dipolar overtones, with straightforward modifications of the following derivations. 

We combined here the dipolar contributions for different B spins within the same molecule, as these contributions cannot be considered statistically independent in the ensemble averaging stage, whereas positions, conformations and orientations of different spin labelled biomolecules can be considered statistically independent. Accordingly, when we average the equation \ref{eq:PDS_0} over the statistical ensemble for the biomolecules distribution in the frozen glassy solvent, each individual background term $b_{s,i}(t)$ can be treated as uncorrelated to all others, and can be averaged independently. We thus get an ensemble averaged PDS signal $V(t)$ in the form

\begin{equation}
    \label{eq:PDS_averaged}
        V(t) = \left<V_s(t)\right> = F(t)\left<\prod_{i=1}^{N}b_{s,i}(t)\right> = F(t)\prod_{i=1}^{N}\left<b_{s,i}(t)\right> = F(t)\prod_{i=1}^{N}b_{i}(t),
\end{equation}
with $F(t) = \left<F_s(t)\right>  = 1-\lambda + \lambda \cdot f(t)$, with $f(t)$ being the normalized modulated part, and $\lambda$ being the modulation depth. Since each biomolecule can assume the same variety of positions, orientations and conformations, we can use the same function $b(t)$ for all individual background contributions, thus arriving at the key relation:

\begin{equation}
    \label{eq:PDS_final}
        V(t) =  F(t)\cdot B(t) = F(t)\left(b(t)\right)^N.
\end{equation}

As a consequence of the equation \ref{eq:PDS_final}, the background functions for two different concentrations $C_1$ and $C_2$ are related to each other as $B(C_1,t) = \left[B(C_2,t)\right]^{C_1/C_2}$. Note that this computation did not use any specific information on the distribution of biomolecules, thus any radial distribution functions, including inhomogeneous cases, can be treated this way.

Practically, it is particularly convenient to treat the case when the two concentrations differ by exactly a factor of 2. Assume we have two mixtures of spin labelled and native type (not labelled) biomolecules with the same total biomolecules concentrations and with the spin dilutions 1:x and 1:2x. The corresponding background functions will be $B_1(t)$ and $B_2(t)$, related to each other as $B_1(t) = \left[B_2(t)\right]^2$. We will accordingly get two different PDS signals $V_1(t)$ and $V_2(t)$, containing the same form factor signal but different intermolecular background contributions. To separate form factor and background contributions, one can use the following relations:

\begin{align}\label{eq:PDS_final_FB}
        B_2(t) = V_1(t)/V_2(t), \notag \\
        F(t) = \left[V_2(t)\right]^2/V_1(t).
\end{align}

\section{Treatment of the spectral diffusion contribution in RIDME}

Differently to constant-time DEER signal, single frequency PDS signals, as for instance RIDME or modulated part of SIFTER (as well as variable-time DEER) have an additional term $B_{SD}(t)$ describing electron spin spectral diffusion within the hyperfine subspectrum. This term does not scale with spin labelled biomolecules concentration, unless the contributing volumes around spin labels start to overlap, which may happen at average electron spin-spin distances of 3-6 nm. Accordingly, equations \ref{eq:PDS_final_FB} do not separate $B_{SD}(t)$ and form factor $F(t)$:  

\begin{align}\label{eq:RIDME_final_FB}
        B_2(t) = V_1(t)/V_2(t), \notag \\
        F(t)\cdot B_{SD}(t) = \left[V_2(t)\right]^2/V_1(t).
\end{align}

The separation of the terms $B_{SD}(t)$ and $F(t)$ can be however accomplished by comparison of the product $V_{\textrm{SD}}(t) = F(t)\cdot B_{SD}(t)$ corresponding to the same mixing time, but measured at two different temperatures. The spectral diffusion term $B_{SD}(t)$ appears due to the nuclear spin diffusion, which is temperature independent in frozen glasses. Contrary to $B_{SD}(t)$, the form factor term $F(t)$ changes upon the change of the ratio between longitudinal relaxation time $T_1$ of the spin labels and the mixing time $T_{\textrm{mix}}$. In a simple approximation the modulation depth in RIDME experiment is proportional to $\left[1 - \exp\left(-T_{\textrm{mix}}/T_1\right)\right]$. However, this accuracy might not suffice and experimental calibration of the modulation depth temperature and mixing time dependence for the given type of spin labels might be necessary. After determining the modulation depth values $\lambda_1$ and $\lambda_2$ corresponding to the given mixing time and the two temperatures $T_1$ and $T_2$, the form factor term can be extracted by first dividing the $V_{\textrm{SD},1}(t)$ and $V_{\textrm{SD},2}(t)$ terms:

\begin{equation}
    \frac{V_{\textrm{SD},2}(t)}{V_{\textrm{SD},1}(t)} = \frac{1-\lambda_2 + \lambda_2\cdot f(t)}{1-\lambda_1 + \lambda_1\cdot f(t)}.
\end{equation}

After making the substitutions $\Delta_1 = \lambda_1/(1-\lambda_1)$, $\Delta_2 = \lambda_2/(1-\lambda_2)$ and $\tilde{V}_{\textrm{SD},12} = [(1-\lambda_1)\cdot V_{\textrm{SD},2}(t)]/[(1-\lambda_2)\cdot V_{\textrm{SD},1}(t)]$, we obtain:

\begin{align}
    \tilde{V}_{\textrm{SD},12} = \frac{1 + \Delta_2 \cdot f(t)}{1 + \Delta_1 \cdot f(t)} \notag \\
    f(t) = \frac{\tilde{V}_{\textrm{SD},12} - 1}{\Delta_2 - \Delta_1 \cdot \tilde{V}_{\textrm{SD},12}}
\end{align}

This allows also determination of the spectral diffusion term from the original product in the equation \ref{eq:RIDME_final_FB} by dividing $\left[V_2(t)\right]^2/V_1(t)$ by the now known form factor $F(t)$.

\section{Treatment of LLPS samples with significant contributions from both phases}

It may happen that the samples after LLPS contain comparable number of spin-labelled molecules in the dispersed and in the condensed state. One can obtain a pure dispersed state sample, if after the LLPS equilibration the liquid biphasic sample at ambient temperature is centrifuged and the supernatant that contains only dispersed phase is taken for freezing and EPR measurements. It is however much harder to make sufficient amount of the condensed phase: due to the much higher biomolecules density in the condensate, filling the whole EPR tube with this phase is rather expensive, albeit still possible. One can, however, use the fact that LLPS leads to a thermodynamic equilibrium of the two new phases, and thus biomolecules concentrations are constant in each phase regardless of its total volume fraction. Also, biomolecules concentration in dispersed phase is typically at few $\mu$M to few tens of $\mu$M levels whereas in the condensate the concentrations of 5-20 mM appear common, i.e. the two concentrations differ by about three orders of magnitude. If we make a second sample with an excess number of biomolecules as compared to the given biphasic sample, all these additional biomolecules will enter the condensed phase. This will lead to substantial relative volume change for the condensed phase but a two-three orders of magnitude smaller relative volume change for the dispersed state. Accordingly, to a good approximation, one can assume the same number of spin labelled biomolecules in the dispersed phases in the two samples, along with accurately known increase of the number of spin labelled biomolecules in the condensed phase in the second sample. Again a simple protocol can be offered if the amounts of biomolecules in the two samples of the same total volume differ by accurately a factor of 2. If we normalize the two PDS signals to have unit amplitude at zero evolution time: $V_n(C)$ and $V_N(2C)$, then a pure condensed-state PDS/EPR signal is obtained by 

\begin{equation}
    V_{\textrm{condensed}} = V_n(2C) - \frac{1}{2}\cdot V_n(C).
\end{equation}

Actually, the same protocol would work as well for many other types of biphasic spectroscopic signals. When applying this protocol one has to be aware of possible differences in the phase memory time of spin labels in the dispersed and in the condensed phase. For instance, at long evolution times in constant-time DEER, there might be an additional magnification factor for slower relaxing EPR signal in dispersed state as compared to the faster relaxing EPR signal in the condensate. In such cases, one has to take care that this additional magnification factor does not win against the ratio of the concentrations in the two phases.

\section{Conclusions and outlook}

The presented protocols for the PDS signals analysis in concentrated and biphasic samples should significantly improve the separability of form factor and background signals, beyond the limitations previously set by standard protocols. For the journal publication we will supplement these analytical derivations by experimental examples on LLPS samples. These measurements are currently ongoing with first results bringing us a confidence for the applicability of the presented derivations.

The work was supported by SNSF grant 205321\_204920 for O.V. and M.Y. and CRSII3\_205922 for L.G.

\bibliography{references} 
\bibliographystyle{rsc} 

\end{document}